\documentclass[12pt]{article}
\usepackage{latexsym}
\usepackage{amsmath}
\usepackage{amssymb}
\catcode`\@=11
\newcount\hour
\newcount\minute
\newtoks\amorpm \hour=\time\divide\hour by 60\minute
=\time{\multiply\hour by 60 \global\advance\minute by-\hour}
\edef\standardtime{{\ifnum\hour<12 \global\amorpm={am}%
        \else\global\amorpm={pm}\advance\hour by-12 \fi
        \ifnum\hour=0 \hour=12 \fi
        \number\hour:\ifnum\minute<10
        0\fi\number\minute\the\amorpm}}
\edef\militarytime{\number\hour:\ifnum\minute<10
0\fi\number\minute}
\def\draftlabel#1{{\@bsphack\if@filesw {\let\thepage\relax
   \xdef\@gtempa{\write\@auxout{\string
      \newlabel{#1}{{\@currentlabel}{\thepage}}}}}\@gtempa
   \if@nobreak \ifvmode\nobreak\fi\fi\fi\@esphack}
        \gdef\@eqnlabel{#1}}
\def\@eqnlabel{}
\def\@vacuum{}
\def\marginnote#1{}
\def\draftmarginnote#1{\marginpar{\raggedright\scriptsize\tt#1}}
\def\draft{
        \pagestyle{plain}
        \overfullrule=2pt
        \oddsidemargin -.1truein
        \def\@oddhead{\sl \phantom{\today\quad\militarytime} \hfil
        \smash{\Large\sl DRAFT} \hfil \today\quad\militarytime}
        \let\@evenhead\@oddhead
        \let\label=\draftlabel
        \let\marginnote=\draftmarginnote
        \def\ps@empty{\let\@mkboth\@gobbletwo
        \def\@oddfoot{\hfil \smash{\Large\sl DRAFT} \hfil}
        \let\@evenfoot\@oddhead}
        \def\@eqnnum{(\theequation)\rlap{\kern\marginparsep\tt\@eqnlabel}%
        \global\let\@eqnlabel\@vacuum}  }

\renewcommand{\theequation}{\thesection.\arabic{equation}}
\renewcommand{\thefootnote}{\fnsymbol{footnote}}
\newcommand{\newsection}{    
\setcounter{equation}{0}\section}
\def\appendix#1{\addtocounter{section}{1}\setcounter{equation}{0}
\renewcommand{\thesection}{\Alph{section}}
\section*{Appendix \thesection\protect\indent \parbox[t]{11.15cm}{#1}}
\addcontentsline{toc}{section}{Appendix \thesection\ \ \ #1}}

\def \la {\label}

\def \b {\beta}

\def\be{\begin{equation}}
\def\ee{\end{equation}}

\def\xx{\mathfrak{ss}}

\catcode`\@=11

\def\bea{\begin{eqnarray}}
\def\eea{\end{eqnarray}}
\def\beann{\begin{eqnarray*}}
\def\eeann{\end{eqnarray*}}
\def\beq{\begin{equation}}
\def\eeq{\end{equation}}
\def\ba{\begin{array}}
\def\ea{\end{array}}
\def\ben{\begin{enumerate}}
\def\een{\end{enumerate}}

 \def \la {\label}
 \def\be{\begin{equation}}
\def\ee{\end{equation}}

\def \la {\label}

\def \ee {\epsilon}

\def \g {\gamma}

\def\a{\alpha }

\def \g {\gamma}

\def \b {\beta}

\def\be{\begin{equation}}
\def\ee{\end{equation}}

\def \la{\label}

\newcommand{\tM}{\text{\tiny $M$}}
\newcommand{\tB}{\text{\tiny $B$}}
\newcommand{\tA}{\text{\tiny $A$}}
\newcommand{\tC}{\text{\tiny $C$}}
\newcommand{\tD}{\text{\tiny $D$}}
\newcommand{\tE}{\text{\tiny $E$}}

\newcommand{\tH}{\text{\tiny $H$}}

\begin{document}
\date{April 2006}
\begin{titlepage}
\begin{center}
\hfill
{}
\vspace{3.5cm}

{\Large \bf  M2-branes, 3-Lie Algebras and Pl\"ucker relations }
\\[.2cm]

\vspace{2.5cm} {\large   G. Papadopoulos}

 \vspace{0.5cm}
Department of Mathematics\\
King's College London\\
Strand\\
London WC2R 2LS, UK\\

\end{center}

\vspace{3.5cm}
\begin{abstract}

We find that the structure constants  4-form of a metric 3-Lie algebra is the sum of the volume forms of orthogonal 4-planes proving  a conjecture
in math/0211170. In particular, there is no metric 3-Lie algebra associated to a $\mathfrak{u}(N)$ Lie algebra for  $N>2$.
We examine the implication of this result on the existence of a multiple M2-brane theory
based on metric 3-Lie algebras.

\end{abstract}
\end{titlepage}
\newpage
\setcounter{page}{1}
\renewcommand{\thefootnote}{\arabic{footnote}}
\setcounter{footnote}{0}

\setcounter{section}{0}
\setcounter{subsection}{0}
\newsection{Introduction}

The Jacobi identity of a metric 3-Lie algebra, $\mathfrak{a}_{[3]}$, is
\bea
F_{\tH[\tA\tB\tC} F^\tH{}_{\tD]\tE\tM}=0~,~~~A,B,C,\dots=0,\dots, n-1
\la{jac}
\eea
where $F$ are the structure constants  and $n$ is the dimension of $\mathfrak{a}_{[3]}$ , respectively.
Compatibility with a metric requires that the structure constants $F$  are skew-symmetric in all indices\footnote{We raise and lower
indices using the compatible metric. Since there is a metric, we do not distinguish between a vector space and its dual.}.
The generalization of (\ref{jac}) to metric $k$-Lie algebras is apparent. In particular for $k=2$,  (\ref{jac}) becomes
the Jacobi identity of a standard metric Lie algebra.

Apart from the interpretation of  (\ref{jac}) as a Jacobi identity for a 3-Lie algebra, it can also be thought of as
a generalized Pl\"ucker relation.   The latter interpretation for the analogous relation for 4-Lie algebras has been instrumental
for  the classification of maximally supersymmetric backgrounds in IIB
supergravity \cite{josegeorge1}. Moreover,
it was conjectured\footnote{In the published version of \cite{josegeorge2} a weaker statement for this conjecture has appeared. The
conjecture in this form can be found in the first version of \cite{josegeorge2}, page 3, see arXiv.} in \cite{josegeorge2} that the only solutions to Jacobi identity are those
for which $F$ is the sum of {\it the volume forms of orthogonal (k+1)-planes}. This was verified
for all metric 3-Lie  algebras and metric 4-Lie  algebras up to and including dimensions 8 and 10, respectively.

More recently, a theory was proposed by Bagger and Lambert  \cite{bl1} for multiple M2-branes
based on metric 3-Lie algebras. Later it was supersymmetrized
by Gustavsson in \cite{gustavsson}, see also \cite{bl2}. This  followed earlier attempts to construct
superconformal ${\cal N}=8$ Chern-Simons
 \cite{schwarz} and multiple M2-brane theories \cite{basu}.  The relation of the multiple M2-branes theory  to the maximally supersymmetric
gauge theory which describes multiple D2-branes has been investigated in
\cite{mp1, lambert, mp2, gran} and some other aspects have been examined
in \cite{berman, raam, morozov, ho, gomis, bergshoeff}. Despite the progress that has been made towards understanding the proposed multiple M2-brane theory based
on metric 3-Lie algebras, there are no known solutions other than those given in \cite{josegeorge2}, see also \cite{schwarz2}.

The main result of this paper is to prove the conjecture of  \cite{josegeorge2} for metric 3-Lie algebras, ie {\it the structure constants of
a metric  3-Lie algebra, $\mathfrak{a}_{[3]}$ with Euclidean signature\footnote{In what follows, we assume
that the metric on $\mathfrak{a}_{[3]}$ is Euclidean unless it is explicitly stated otherwise.},  can be written as
\bea
F=\sum_r\mu^r\,  d{\rm vol}(V_r)~,~~~~ V_r\subset \mathfrak{a}_{[3]}
\la{main}
\eea
where the 4-planes $V_r$ and $V_{r'}$ are mutually orthogonal  for $r\not= r'$ and $\mu^r$ are constants}.
This conjecture for metric 3-Lie algebras has also appeared more recently in \cite{gustavsson, schwarz2, ho}.

To prove our result, first observe that
given a vector $X$ in $\mathfrak{a}_{[3]}$, one can associate a metric Lie algebra  $\mathfrak{a}_{[2]}(X)$ to  $\mathfrak{a}_{[3]}$ defined
as the orthogonal complement of $X$ in $\mathfrak{a}_{[3]}$ with structure constants $i_XF$. It is easy to verify
that $i_XF$ satisfy the Jacobi identity using (\ref{jac}). Then we demonstrate the following two statements:
\begin{itemize}
\item If $\mathfrak{a}_{[3]}$ admits  an associated Lie algebra $\mathfrak{a}_{[2]}(X)$ and
$\mathfrak{a}_{[2]}(X)$  does not have
a bi-invariant 4-form, then $F$ is volume form of a  4-plane.

\item If all the associated metric Lie algebras of $\mathfrak{a}_{[3]}$ are
 $\mathfrak{a}_{[2]}(X)=\oplus^\ell\mathfrak{u}(1)\oplus \mathfrak{\xx}$, for some
$\ell\geq 0$, where $\mathfrak{\xx}$ is a semi-simple Lie algebra which commutes with $\oplus^\ell\mathfrak{u}(1)$, then $F$ is as in (\ref{main}).

\end{itemize}
We complete the proof by showing that all metric Lie algebras are isomorphic to $\oplus^\ell\mathfrak{u}(1)\oplus \mathfrak{\xx}$ in appendix A.

This paper is organized as follows. In section two, we give the proof of the first statement. In section 3,
we give the proof of the second statement and in section 4, we examine the applications of our results in the
context of multiple M2-branes. In appendix A, we demonstrate that all metric Lie algebras are isomorphic
to $\oplus^\ell\mathfrak{u}(1)\oplus \mathfrak{\xx}$.

\newsection{3-Lie algebras and invariant 4-forms}

The Jacobi identity of metric $k$-Lie algebras is an over-constrained  quadratic equation for the structure constants $F$. In particular, it
is easy to see that the ratio of the number of relations given by the Jacobi identity to the number of components of the structure constants $F$
grows as $n^{k-1}$ for large $n$. In particular, it grows linearly for Lie algebras and quadratically for 3-Lie algebras.
So it is clear that the structure constants of $k$-Lie algebras for $k>2$ are more restricted than those
of standard Lie algebras.

To prove the first statement in the introduction, without loss of generality, take   the vector field $X$ to be along the $0$ direction. Then
   split the indices as $A=(0, i)$, $B=(0,j)$ and so on, with $i,j, \dots=1, \dots, n-1$. Setting $A=M=0$ and the rest of the
free indices in the range $1, \dots, n-1$ in (\ref{jac}), it is easy to see that
\bea
f_{ijk}=F_{0ijk}
\eea
satisfy the Jacobi identity of standard Lie algebras and $f$ are the structure constants of  $\mathfrak{a}_{[2]}(X)$. Thus
we have written $F$ as
\bea
F={1\over 3!} f_{ijk}\,\,  e^0\wedge  e^i \wedge e^j \wedge e^k+ {1\over 4!} \phi_{ijkl}\, e^i \wedge e^j \wedge e^k\wedge e^l
\la{0phi}
\eea
where $(e^0, e^i)$, $i=1,\dots, n-1$, is an orthonormal basis.

Next set $M=0$ and the rest of the free indices in the range $1, \dots, n-1$ in (\ref{jac}). Using the skew-symmetry of $F$, one finds that
\bea
\phi_{h[ijk} f^h{}_{d]e}=0~.
\la{invphi}
\eea
This implies that the 4-form $\phi$ is bi-invariant with respect to $\mathfrak{a}_{[2]}(X)$. Since by assumption such form cannot exist,
we conclude that $\phi=0$. Thus, we find so far that
\bea
F={1\over 3!}f_{ijk} e^0\wedge  e^i \wedge e^j \wedge e^k~.
\eea

Next taking all free indices in (\ref{jac}) in the range $1,\dots, n-1$, we find that
\bea
f_{[ijk} f_{h]de}=0~.
\la{stplu}
\eea
This is the classical Pl\"ucker relation which implies that $f$ is a simple\footnote{The term ``simple'' is used in two different ways. In the context of
Lie algebras it means the usual simple Lie algebras like $\mathfrak{su}$ and $\mathfrak{so}$. In the context
of forms it is used in the Pl\"ucker sense, ie a $p$-form is simple iff it is the wedge product of $p$ 1-forms, see eg \cite{alg}.} form. Thus
one  concludes that the only solution to (\ref{jac}) is
\bea
F=\mu\, e^0\wedge e^1\wedge e^2\wedge e^3~,
\la{sim}
\eea
for some constant $\mu$, where we have chosen the four 1-forms, without loss of generality, to lie in the first four directions.
This proves the first statement of the paper.

The assumption that  $\mathfrak{a}_{[2]}(X)$ does not admit a bi-invariant 4-form is not very strong. To see this,
it is known that  bi-invariant forms on the Lie algebra of a group give rise to parallel forms with respect to the Levi-Civita
connection on the associated simply connected
group manifold. So if $\mathfrak{a}_{[2]}(X)$ admits a bi-invariant 4-form, the associated group manifold $G$ admits
a parallel 4-form which is necessarily harmonic. So for compact
Lie groups, parallel 4-forms represent  non-trivial classes in the 4-th deRham cohomology of $G$.
Thus if a compact Lie group admits a parallel 4-form, then $H_{dR}^4(G)\not=0$. However for a large class
of Lie groups, which includes all semi-simple ones,  $H_{dR}^4(G)=0$. Thus we conclude that if an associated
Lie algebra $\mathfrak{a}_{[2]}(X)$ to $\mathfrak{a}_{[3]}$ is semi-simple, then the assumption
of the first statement in the introduction is satisfied and the only solution is as in (\ref{sim}).

\newsection{3-Lie algebras and  Lie algebras}

To proceed to prove the second statement, first observe that we have already demonstrated the result for $\ell=0$ at the end of the previous
section. Next,
 let us consider the case that
for some $X$ the associated Lie algebra is
\bea
\mathfrak{a}_{[2]}(X)=\mathfrak{u}(1)\oplus \mathfrak{\xx}~.
\eea
In such as case,  $H_{dR}^4(G)\not=0$ and there are bi-invariant 4-forms on $\mathfrak{a}_{[2]}(X)$.
Such bi-invariant 4-forms  are of the type
\bea
\phi=e^{n-1}\wedge \varphi= {1\over 3!} \varphi_{\a\b\g}\,\, e^{n-1}\wedge e^\a\wedge e^\b \wedge e^\g~,~~~\a,\b,\g=1,\dots, n-2~,
\eea
where $\varphi$ is a bi-invariant form on the semi-simple part $\mathfrak{\xx}$ of $\mathfrak{a}_{[2]}(X)$.
Without loss of generality, we have chosen the $\mathfrak{u}(1)$ direction along  $e^{n-1}$.
Thus so far we have
\bea
F={1\over 3!} f_{\a\b\g} e^0\wedge e^\a\wedge e^\b \wedge e^\g+ {1\over 3!} \phi_{\a\b\g}\,\, e^{n-1}\wedge e^\a\wedge e^\b \wedge e^\g~.
\eea
Since $\varphi$ is a bi-invariant 3-form on a semi-simple Lie algebra, $\varphi$ is a linear
combination of the structure constants $f_r$ of the simple Lie algebras\footnote{Every semi-simple Lie algebra
decomposes into an orthogonal sum of simple Lie algebras, $\mathfrak{\xx}=\oplus^r \mathfrak{s}_r$,  with respect to the
bi-invariant inner product,
such that  $[\mathfrak{s}_r,\mathfrak{s}_{r'}]=0$
for $r\not=r'$.} in $\mathfrak{\xx}=\oplus^r \mathfrak{s}_r$. Thus $F$ can be rewritten as
\bea
F=\sum_r (\mu^r e^0+\nu^r e^{n-1})\wedge f_r~,
\eea
where $\mu^r\not=0$ and $\nu^r$ are some constants.

Next using that $f_r$ and $f_{r'}$,  for $r\not= r'$, are mutually orthogonal, the Jacobi identity (\ref{jac}) implies that
$\mu^r e^0+\nu^r e^{n-1}$ and $\mu^{r'} e^0+\nu^{r'} e^{n-1}$ are mutually orthogonal as well. This can be easily seen by setting
$A,B,C$ to take values in the $\mathfrak{s}_r$  and $D, E, M$ to take values in the $\mathfrak{s}_{r'}$ component of $\mathfrak{\xx}$, respectively. Moreover applying the Jacobi
(\ref{jac}) for the same simple component, ie allowing all $A, B, C, D, E, M$ to take values in the same simple component $\mathfrak{s}_r$,
one finds that $f_r$
satisfies the standard Pl\"ucker relation, as in (\ref{stplu}),  and so $f_r$  must be a simple 3-form.
In conclusion, $F$ is the sum of volume forms of at most two orthogonal 4-planes, ie without loss of generality, it can be written as
\bea
F=\mu\, e^0\wedge e^1\wedge e^2\wedge e^3+\nu\, e^4\wedge e^5\wedge e^6\wedge e^7~,
\eea
where $\mu$ and $\nu$ are constants.

This can be  extended to  metric 3-Lie algebras for which all associated Lie algebras are
\bea
\mathfrak{a}_{[2]}(X)=\oplus^\ell\mathfrak{u}(1)\oplus \mathfrak{\xx}~,~~~ \ell> 1~.
\eea
 To begin  write $F=e^0\wedge f+\phi$ as in (\ref{0phi}), where $f$ are the structure constants of $\mathfrak{\xx}$
and $\phi$ satisfies (\ref{invphi}).
Since semi-simple Lie algebras do not admit bi-invariant 1-,  2- and 4-forms, see eg \cite{top}, and $\oplus^\ell\mathfrak{u}(1)$
commutes with $\mathfrak{\xx}$, the most general
invariant 4-form $\phi$  which satisfies (\ref{invphi}) is
\bea
\phi=\sum_{I} \rho^I\wedge \varphi_I+\xi~,~~~
\eea
where $\rho^I$ are the 1-forms along the $\oplus^\ell\mathfrak{u}(1)$ directions,  $\varphi_I$ are bi-invariant 3-forms of $\mathfrak{\xx}$ and $\xi$
is a 4-form along the $\oplus^\ell\mathfrak{u}(1)$ directions.

Since the bi-invariant 3-forms of semi-simple Lie algebras are linear combinations of those associated with the structure
constants of the simple components $\mathfrak{s}_r$, we have
\bea
\varphi_I= \nu_I{}^r f_r~.
\eea
So $F$ can be rewritten as
\bea
F=\sum_{r} \sigma^r\wedge f_r+ \xi~,
\eea
for some constants $\mu^r\not=0$ and $\nu_I{}^r$, where
\bea
\sigma^r=  \mu^r e^0+ \sum_I \nu_I{}^r \rho^I~.
\eea
Using that $f_r$ and $f_{r'}$,  for $r\not= r'$, are mutually orthogonal, the Jacobi identity (\ref{jac}) implies that
$\sigma^r$ and $\sigma^{r'}$ are mutually orthogonal as well. Thus there is an orthogonal transformation in $\oplus^\ell \mathfrak{u}(1)$
such that $F$ can be written, without loss of generality, as
\bea
F=\sum_{r} \lambda_r\,\, e^r\wedge f_r+\xi~,
\eea
for some constants $\lambda_r$,
where $e^r$ belong to an orthonormal basis in the $\oplus^\ell \mathfrak{u}(1)$ directions, ie in particular $e^r\perp e^{r'}$ for $r\not=r'$ and
 $i_{e^r} f_s=0$ for all $r$ and $s$.
As in previous cases,  allowing all indices $A, B, C, D, E, M$ of (\ref{jac}) to take values in the same simple component $\mathfrak{s}_r$,
one finds that $f_r$
satisfies the standard Pl\"ucker relation,  (\ref{stplu}),  and so $f_r$  must be a simple 3-form. Thus the
component of $F$ {\it orthogonal} to $\xi$ is as in (\ref{main}).

Furthermore, using the orthogonality of $f_r$ and $\xi$ and the Jacobi identity (\ref{jac}), one finds that
\bea
i_{e^r}\xi=0~.
\la{fin}
\eea
This can be easily seen from (\ref{jac}) by setting
$A,B,C$ to take values in the $\mathfrak{s}_r$  and $D, E, M$ to take values in the $\oplus^\ell \mathfrak{u}(1)$.
A consequence of (\ref{fin}) is that the 4-form $\xi$ on $\oplus^\ell \mathfrak{u}(1)$ satisfies (\ref{jac}),
ie {\it $\oplus^\ell \mathfrak{u}(1)$ is also a metric 3-Lie algebra, $ \mathfrak{b}_{[3]}$,  with structure constants $\xi$}.
Since the dimension of $ \mathfrak{b}_{[3]}\subset \mathfrak{a}_{[3]}$  is strictly less than that of the original metric
3-Lie algebra $\mathfrak{a}_{[3]}$, the analysis can be repeated and it will terminate after a finite number of steps.
To summarize, we have demonstrated
(\ref{main}) under the assumption that all the associated Lie algebras $\mathfrak{a}_{[2]}(X)$ of $\mathfrak{a}_{[3]}$ are isomorphic to
$\oplus^\ell\mathfrak{u}(1)\oplus \mathfrak{\xx}$.

To prove the conjecture in \cite{josegeorge2} for metric 3-Lie algebras, it remains to show that all metric Lie algebras are isomorphic to
$\oplus^\ell\mathfrak{u}(1)\oplus \mathfrak{\xx}$. This is done in appendix A and so we establish (\ref{main}).

\newsection{Multiple M2-branes}

Some of the results we have presented for metric 3-Lie algebras can be extended to metric k-Lie algebras for $k>3$. This together with the
$n^{k-1}$ growth of relations on the structure constants mentioned in section 2,
strengthens the conjecture in \cite{josegeorge2}. In particular,  it looks likely that
the structure constants of metric k-Lie algebras are sums of  volume forms of orthogonal   $(k+1)$-planes. Though there is not a  proof for this
for $k>3$. Our result also relies on the existence of a Euclidean inner product on the 3-Lie algebras. Thus the Lorentzian
signature case has to be examined separately. As it has already been mentioned in \cite{josegeorge2}, it is unlikely that (\ref{main})
holds for other signatures.

A  consistency condition for the validity of a multiple M2-brane theory for $N\geq 2$ is the derivation from it  of the maximally
supersymmetric gauge theory in 3-dimensions which describes $N$ coincident  D2-branes. The gauge group of the latter for D2-branes in flat
space is $U(N)$. Our result demonstrates that the expected  $U(N)$ gauge group cannot be recovered from a
metric 3-Lie algebra for $N>2$. This in particular includes the identification
of  the gauge group $U(N)$  in a metric 3-Lie algebra as in eg \cite{mp1, lambert, mp2}. So
 the  consistency check on the multiple M2-brane theory  with that of  D2-branes cannot  be met for $N>2$. In fact our result excludes all other
possibilities for other gauge groups apart from $\times^\ell SU(2)$.

The above results clearly indicate that some of the assumptions
must be weakened. This has already been anticipated by Gran, Nilsson and Petersson in \cite{gran} where they considered a theory of
multiple M2-branes using an $F$ which is not skew-symmetric. From this theory one can consistently
derive the effective theory of $N$ coincident  D2-branes. However, this multiple M2-brane theory is not described
by an action. The dynamics of the theory is given  in terms of field equations. Though one may be able to define
a  Hamiltonian and so quantize the theory.
Alternatively, one can consider multiple M2-brane theories based on Lorentzian metric 3-Lie algebras. Again, it is not apparent that such
a choice will lead to a consistent theory. First, the possibility remains
open that $F$ may again be written as (\ref{main}). Some  partial results in \cite{josegeorge2} indicate this. In addition, the Hamiltonian
of the theory will  be unbounded from below  which in turn may lead to difficulties with the particle physics interpretation of the theory.

\section*{Acknowledgements}
 I would like to thank U Gran for many helpful discussions and comments.

 \setcounter{section}{0}

\appendix {Metric Lie algebras}

 It is known that  any Lie algebra $\mathfrak{g}$, which is not semi-simple,  contains an {\it invariant (normal) solvable
subalgebra}, $\mathfrak{r}$,  called the {\it radical},  such that $\mathfrak{g}/\mathfrak{r} $ is semi-simple, see eg  \cite{gilmore} page 234.
This follows from the definition of semi-simple Lie algebras.
Now assume that $\mathfrak{g}$  admits a Euclidean invariant metric $B$. In such case,  one can define
the semi-simple algebra $\mathfrak{\xx}$  as the
{\it orthogonal complement} of $\mathfrak{r}$ in $\mathfrak{g}$.
Using this and the invariance property of $B$ which schematically can be written as
\bea
B([\mathfrak{g}, \mathfrak{g}], \mathfrak{g})+ B(\mathfrak{g}, [\mathfrak{g},\mathfrak{g}])=0~,
\eea
 one finds that
\bea
[\mathfrak{\xx}, \mathfrak{\xx}]\subseteq \mathfrak{\xx}~,~~~[\mathfrak{\xx}, \mathfrak{r}]\subseteq \mathfrak{r}~,~~~
[\mathfrak{r}, \mathfrak{r}]\subseteq \mathfrak{r}~.
\eea
So to prove the statement, we have first to show that for metric Lie algebras $\mathfrak{g}$, $\mathfrak{r}$ is abelian.
From the definition of solvable algebras there is an $n$ such that $\mathfrak{r}^{n}$ is an abelian Lie algebra,
where $\mathfrak{r}^{i+1}=[\mathfrak{r}^{i}, \mathfrak{r}^{i}]$ and $\mathfrak{r}^{0}=\mathfrak{r}$.
 The metric $B$ restricted on $\mathfrak{r}$ and $\mathfrak{r}^{n}$ is not degenerate. Since $\mathfrak{r}^{n}$ abelian, it is easy to see that
 \bea
 B([\mathfrak{r}^{n}, \mathfrak{r}^{n}],\mathfrak{r}^{n-1})+B(\mathfrak{r}^{n}, [\mathfrak{r}^{n},\mathfrak{r}^{n-1}]) =
B(\mathfrak{r}^{n}, [\mathfrak{r}^{n},\mathfrak{r}^{n-1}])=0~.
 \eea
So if $[\mathfrak{r}^{n-1},\mathfrak{r}^{n}]\not=\{0\}$, it is orthogonal to the whole
 of $\mathfrak{r}^{n}$. But
  $[\mathfrak{r}^{n-1},\mathfrak{r}^{n}]\subseteq \mathfrak{r}^{n}$
 and since the metric is not degenerate, one concludes that $[\mathfrak{r}^{n-1},\mathfrak{r}^{n}]=0$.
Similarly
\bea
 B([\mathfrak{r}^{n-1}, \mathfrak{r}^{n}],\mathfrak{r}^{n-1})+B(\mathfrak{r}^{n}, [\mathfrak{r}^{n-1},\mathfrak{r}^{n-1}]) =
B(\mathfrak{r}^{n}, [\mathfrak{r}^{n-1},\mathfrak{r}^{n-1}])=0~.
 \eea
But $[\mathfrak{r}^{n-1},\mathfrak{r}^{n-1}]= \mathfrak{r}^{n}\not=\{0\}$.  Since the metric is not degenerate, one concludes that
  $[\mathfrak{r}^{n-1},\mathfrak{r}^{n-1}]=0$. Therefore $\mathfrak{r}^{n-1}$
is also abelian and so  $\mathfrak{r}^{n-1}=\mathfrak{r}^{n}$. Continuing in this way,
one can show that $\mathfrak{r}=\mathfrak{r}^{n}$ is abelian. So without loss of generality,
we  can write $\mathfrak{r}=\oplus^\ell\mathfrak{u}(1)$.

It remains to show that $[\mathfrak{\xx}, \mathfrak{r}]=0$. This follows again from the invariance of the metric. Indeed
\bea
B([\mathfrak{r},\mathfrak{\xx}], \mathfrak{r})+B(\mathfrak{\xx}, [\mathfrak{r}, \mathfrak{r}])=
B([\mathfrak{r},\mathfrak{\xx}], \mathfrak{r})=0~,
\eea
using that $\mathfrak{r}$ is abelian.
Thus if   $[\mathfrak{r},\mathfrak{\xx}]\not=\{0\}$, the subspace  $[\mathfrak{r},\mathfrak{\xx}]$ of $\mathfrak{r}$
is orthogonal to the whole $\mathfrak{r}$. Since the metric is non-degenerate, one again concludes that  $[\mathfrak{r},\mathfrak{\xx}]=0$.
Thus all metric Lie algebras can be written as $\oplus^\ell \mathfrak{u}(1)\oplus \mathfrak{\xx}$ with
$\mathfrak{\xx}$ to commute with $\oplus^\ell \mathfrak{u}(1)$.

\setcounter{section}{0}

\end{document}